The Perturbed NLS Equation and Asymptotic Integrability

Yair Zarmi[1,2]
Ben-Gurion University of the Negev, Israel
[1]Department of Energy & Environmental Physics
Jacob Blaustein Institutes for Desert Research, Sede-Boqer Campus, 84990
[2]Department of Physics, Beer-Sheva, 84105

ABSTRACT

The perturbed NLS equation is asymptotically integrable if the zero-order approximation to the solution is a single soliton. It is not integrable for any other zero-order solution. In the standard application of a Normal Form expansion, integrability is recovered if, from second order onwards, the perturbation obeys algebraic constraints. In general, these are not satisfied. As a result, the Normal Form is not asymptotically integrable; its solution, the zero-order approximation, loses the simple structure of the solution of the unperturbed equation. This work exploits the freedom in the perturbative expansion to propose an alternative, for which the loss of integrability only affects higher-order terms in the expansion of the solution. The Normal Form remains asymptotically integrable, and the zero-order approximation, not affected by the loss of integrability, has the same structure as the solution of the unperturbed equation. In the case of a multiple-soliton zero-order approximation, the loss of integrability affects the higher-order corrections to the solution in a finite domain around the origin in the *x*-*t* plane, the soliton interaction region. The effect decays exponentially in all directions in the plane as the distance from the origin grows. The computation is carried through second order.

## I. Introduction

The NLS equation is integrable: It has an infinite hierarchy of symmetries [1-9]; its single- and multiple-soliton solutions can be obtained in closed form [10-14], and obey an infinite sequence of conservation laws [1-3]. When the terms that have been neglected in the derivation of the NLS equation are added as a small perturbation, the solution of the perturbed equation is analyzed by a version of the Normal Form expansion [15-17], applied to perturbed PDE's: the KdV equation in [18-22], the Burgers equation in [23, 24] and the NLS equation in [6, 25, 26]. The zero-order approximation to the full solution, $\varphi$, is the solution of Normal From (NF). In this standard application of the method, the higher-order corrections in the Near Identity Transformation (NIT) are differential polynomials in $\varphi$. For a general zero-order solution, the three perturbed equations mentioned above lose integrability [6, 18-26]. The only exception is the case when $\varphi$ is a single-wave (front or soliton) solution with the same the structure as the unperturbed solution. In the latter case, the NF and $\varphi$ satisfy the integrability properties of the unperturbed equation, and the effect of the NF is a mere updating of the wave parameters. The higher-order corrections are, indeed, differential polynomials in $\varphi$, and are bounded. The effect of the loss of integrability for a general $\varphi$ depends on the expansion algorithm. In the standard analysis, none of the properties defining integrability is retained from some order and onwards, unless specific combinations of numerical coefficients that appear in the perturbation vanish [6, 18-26]. Usually, these combinations, the *algebraic obstacles to integrability*, do not vanish for physical systems. The NF is then not integrable; its solution, $\varphi$, does not have the simple multiple-wave structure of the solution of the unperturbed equation.

This work focuses on the case of the perturbed NLS equation, where the loss of integrability shows up starting from second order [6, 26]. A way is proposed to circumvent the algebraic impasse and the resulting loss of integrability of the NF in the standard NF analysis through second order. The assumption that, starting from second order, the terms in the NIT are differential polynomials in the zero-order approximation, is abandoned, and non-polynomial dependence of the higher-order corrections on the zero-order solution, $\varphi$, is allowed [1]. The work focuses on *dynamical obstacles*. These are differential polynomials contained in the perturbation, which are responsible for the loss of integrability for the full, perturbed equation.

---

[1] The problem arises already in the NF analysis of perturbed ODE's, where, customarily, higher-order corrections in the NIT are assumed to be polynomials in the zero-order term. This assumption works for systems of *autonomous* equations with a *linear unperturbed part*. Inconsistencies may emerge in all other cases, and are resolved by allowing for explicit time dependence in the NIT. The issue is discussed in part in [27]

In the absence of guidance regarding the non-polynomial dependence on $\varphi$ in the higher-order corrections to the solution, explicit $t$- and $x$-dependence is incorporated in them, above any differential polynomial allowed by the formalism. The added freedom enables one to shift the effect of the obstacles to integrability from the NF to the NIT. The NF then remains integrable. Its solution ($\varphi$, the zero-order approximation) retains the multiple-soliton character of the solution of the unperturbed equation. Moreover, as in the single-soliton case, the NF merely updates the parameters of each soliton by the effect of the perturbation. In addition, the hierarchy of symmetries of the unperturbed equation is also the hierarchy of symmetries for the NF, and $\varphi$ obeys the same infinite sequence of conservation laws, obeyed by the unperturbed solution.

The new expansion algorithm yields a "canonical" second-order dynamical obstacle, expressible in terms of symmetries of the NLS equation. This obstacle vanishes identically for a single-soliton. (Obstacles that are not in the canonical form need not vanish for a single-soliton. That they do not emerge in the latter case is discovered through explicit calculation.) For a multiple-soliton solution, the canonical obstacle does not vanish, but decays exponentially as the distance from the soliton-interaction region (a finite domain around the origin in the $x$-$t$ plane) grows. The only effect of the loss of integrability is a bounded, exponentially decaying, non-polynomial term in the second-order correction to the full solution. Hence, the notion of asymptotic integrability of the perturbed NLS equation obtains a new meaning: The NF, together with its solution, the zero-order approximation, remains rigorously integrable, and the effect of the loss of integrability on the higher-order corrections decays as the distance from the origin grows. The conclusions of the analysis presented here apply also when the perturbation added to the NLS equation is not small, provided the wave numbers of all the solitons are sufficiently small. Similar results have been obtained in the case of the perturbed KdV equation [28-30].

## 2. NLS equation

### 2.1 Solutions

The NLS equation,

$$\varphi_t = F_0[\varphi] \equiv i\,\varphi_{xx} + 2i|\varphi|^2\varphi\,, \tag{2.1}$$

is integrable. Its single soliton solution is given by [10-14]

$$\varphi(t,x) = k\frac{e^{i\gamma}}{\cosh\beta} \quad . \tag{2.2}$$

In Eq. (2.2),

$$\begin{aligned} \beta &= k\left(x - x_0 + v\left(t - t_0\right)\right) \\ \gamma &= \omega\left(t - t_0\right) + V\left(x - x_0\right) \\ \omega &= k^2 - \left(v^2/4\right) \;,\; V = -\left(v/2\right) \end{aligned} \quad . \tag{2.3}$$

Of the multiple-soliton solutions, the two-soliton solution is addressed in this work [10-14, 26]:

$$\varphi(t,x) = \frac{1}{D}\left\{ \begin{aligned} &\frac{k_1 e^{i\gamma_1}}{\cosh\beta_1}\left[\left(k_1\right)^2 - \left(k_2\right)^2 + \frac{\left(v_1 - v_2\right)^2}{4} + i\,k_2\left(v_2 - v_1\right)\tanh\beta_2\right] + \\ &\frac{k_2 e^{i\gamma_2}}{\cosh\beta_2}\left[\left(k_2\right)^2 - \left(k_1\right)^2 + \frac{\left(v_1 - v_2\right)^2}{4} + i\,k_1\left(v_1 - v_2\right)\tanh\beta_1\right] \end{aligned} \right\} \quad . \tag{2.4}$$

In Eq. (2.4), $\beta_i$ and $\gamma_i$ are defined for $i$ =1, 2 as in Eq. (2.3), and

$$D = \left(k_1 - k_2\right)^2 + \frac{\left(v_1 - v_2\right)^2}{4} + 2\,k_1\,k_2\,\frac{\cosh\left[\beta_1 - \beta_2\right] - \cos\left[\gamma_1 - \gamma_2\right]}{\cosh\beta_1\cosh\beta_2} \quad . \tag{2.5}$$

**2.2 Symmetries**

Eq. (2.1) has an infinite hierarchy of *symmetries*, which are solutions of its linearization [1-8]:

$$\partial_t S_n = \frac{\partial}{\partial\nu}F_0\left[\varphi + \nu\,S_n\right]\Big|_{\nu=0} = i\,\partial_x{}^2 S_n + 2i\left(2\varphi\varphi^* S_n + \varphi^2 S_n{}^*\right) \quad . \tag{2.6}$$

The first five symmetries, required for a second-order perturbative analysis, are:

$$\begin{aligned} S_0 &= i\varphi \\ S_1 &= \partial_x\varphi \\ S_2 &= i\varphi_{xx} + 2i|\varphi|^2\varphi \\ S_3 &= \varphi_{xxx} + 6|\varphi|^2\varphi_x \\ S_4 &= i\varphi_{xxxx} + 2i\varphi^2\varphi^*{}_{xx} + 8i|\varphi|^2\varphi_{xx} + 6i\varphi^*\left(\partial_x\varphi\right)^2 + 4i|\varphi_x|^2\varphi + 6i|\varphi|^4\varphi \end{aligned} \quad . \tag{2.7}$$

Replacing the r.h.s. of Eq. (2.1) by any symmetry, yields an equation that has the same single- and multiple- soliton solutions as of Eq. (2.1), with known modifications of the dependence of the phases, $\gamma_i$, on $k_i$ and $v_i$. In addition, the symmetries are resonant terms that have the capacity to generate unbounded terms in the expansion of the solution of the perturbed NLS equation, and are the building blocks of an integrable NF.

**2.3 Conservation laws**

The solutions of Eq. (2.1), which vanish sufficiently rapidly at $x = \pm \infty$ (soliton solutions included), obey an infinite sequence of conservation laws [1-3]. An infinite sequence of densities, $\rho_n[\varphi]$ can be constructed, for which the quantities

$$I_n = \int_{-\infty}^{+\infty} \rho_n[\varphi] dx \tag{2.8}$$

are constants of motion. The first three densities are [1-3, 26]:

$$\begin{aligned} \rho_0 &= |\varphi|^2 \\ \rho_1 &= i\varphi\varphi^*_x \\ \rho_2 &= |\varphi|^4 - |\varphi_x|^2 \end{aligned} \quad . \tag{2.9}$$

**3. Standard NF analysis of the perturbed NLS equation [6, 25, 26]**

In this Section, the analysis of the perturbed NLS equation and the effects of obstacles to integrability in the standard NF expansion are reviewed. Through second order, the equation reads:

$$\begin{aligned} &\psi_t = F[\psi] = \\ &\quad i\psi_{xx} + 2i|\psi|^2\psi \\ &\quad + \varepsilon\left(\alpha_1\psi_{xxx} + \alpha_2|\psi|^2\psi_x + \alpha_3\psi^2\psi_x^*\right) \\ &\quad + \varepsilon^2 i\begin{pmatrix} \beta_1\psi_{xxxx} + \beta_2|\psi|^2\psi_{xx} + \beta_3\psi^*\left(\psi_x^2\right) \\ + \beta_4\psi^2\psi_{xx}^* + \beta_5\psi|\psi_x|^2 + \beta_6|\psi|^4\psi \end{pmatrix} + O\left(\varepsilon^3\right) \end{aligned} \quad . \tag{3.1}$$

One expands $\psi$ in a *Near Identity Transformation* (NIT):

$$\psi = \varphi + \varepsilon\varphi^{(1)} + \varepsilon^2\varphi^{(2)} + O\left(\varepsilon^3\right) \quad . \tag{3.2}$$

The evolution of the zero-order term, $\varphi(t,x)$, is expected to be governed by the *Normal Form* (NF), which is constructed from, the symmetries of the unperturbed equation:

$$\varphi_t = S_2[\varphi] + \varepsilon\alpha_1 S_3[\varphi] + \varepsilon^2 \beta_1 S_4[\varphi] + O(\varepsilon^3). \tag{3.3}$$

Eq. (3.3) is integrable; it is also solved by the single- and multiple-soliton solutions of Eq. (2.1), with the following updating of parameters for each soliton, found by direct computation:

$$\begin{aligned} \omega &= k^2 - (v^2/4) - \varepsilon\alpha_1\left\{v\left(k^2 + (v^2/4)\right)\right\} + \varepsilon^2\left\{\alpha_1^2\left(\tfrac{5}{4}k^4 - \tfrac{3}{8}k^2v^2 - \tfrac{27}{64}v^4\right) + \beta_1\left(k^4 - \tfrac{1}{2}k^2v^2 - \tfrac{3}{16}v^4\right)\right\} \\ V &= -(v/2) - \varepsilon\alpha_1\left\{\tfrac{1}{2}k^2 - \tfrac{3}{8}v^2\right\} + \varepsilon^2 v\left\{\alpha_1^2\left(\tfrac{3}{4}k^2 - \tfrac{9}{16}v^2\right) + \beta_1\left(k^2 - \tfrac{1}{4}v^2\right)\right\} \end{aligned} . \tag{3.4}$$

**3.1 Loss of integrability of NF**

In general, it is impossible to obtain Eq. (3.3) in the standard analysis. The most general differential polynomials allowed by the formalism for the $\varphi^{(1)}$ and $\varphi^{(2)}$ in Eq. (3.2) are:

$$\begin{aligned} &\varphi^{(1)} = ia\varphi_x + ibq\varphi \\ &\varphi^{(2)} = A\varphi_{xx} + B\varphi^2\varphi^* + Cp\varphi + Dq\varphi_x + Eq^2\varphi \\ &\left(q = \partial_x^{-1}|\varphi|^2,\ p = \partial_x^{-1}\left(\varphi\varphi^*_x\right)\right) \end{aligned} . \tag{3.5}$$

The values of the coefficients in $\varphi^{(1)}$ are readily found to be:

$$\begin{aligned} a &= -\tfrac{3}{2}\alpha_1 + \tfrac{1}{4}\alpha_2 - \tfrac{1}{4}\alpha_3 \\ b &= -3\alpha_1 + \tfrac{1}{2}\alpha_2 \end{aligned} . \tag{3.6}$$

However, in general, the coefficients in $\varphi^{(2)}$ can be determined only if [26]:

$$\begin{aligned} \Gamma \equiv\ & 18\alpha_1^2 - 3\alpha_1\alpha_2 + \alpha_2\alpha_3 - 2\alpha_3^2 \\ & + 24\beta_1 - 2\beta_2 - 4\beta_3 - 8\beta_4 + 2\beta_5 + 4\beta_6 = 0 \end{aligned} . \tag{3.7}$$

The reason is that $\varphi^{(2)}$ has five independent terms, whereas the second-order perturbation has six.

In general, physical systems do not obey Eq. (3.7). As a result, the differential polynomial assumed for $\varphi^{(2)}$ cannot account for all the terms generated by the perturbative procedure in second order. The only way out in the standard analysis is to assign the unaccounted terms to the NF, Eq. (3.3), which is then modified into:

$$\varphi_t = S_2[\varphi] + \varepsilon\alpha_1 S_3[\varphi] + \varepsilon^2\left(\beta_1 S_4[\varphi] + \Gamma\tilde{R}[\varphi]\right) + O\left(\varepsilon^3\right) \quad . \tag{3.8}$$

$\tilde{R}[\varphi]$ is the dynamical obstacle, representing all the terms that have remained unaccounted for.

The structure of the obstacle in Eq. (3.8) is not unique because the coefficients in $\varphi^{(2)}$ are not determinable: Different choices of the coefficients generate different unaccounted terms that one is forced to include in the NF. What is common to these terms is that they are not symmetries. Hence, their addition to the NF spoils its integrability [26]. This results in a zero-order solution that does not have the simple multiple-soliton structure of the unperturbed solution.

The only exception is the case of a single-soliton zero-order solution, $\varphi$. In this case, the coefficients in $\varphi^{(2)}$ except for one, which remains free, can be determined. Their values are given in the Appendix (Eq. (A.1)), where *B* has been chosen as the free coefficient.

### 3.2 Loss of symmetries

In general, it is impossible to construct symmetries for the perturbed NLS equation. As a substitute, the concept of approximate symmetries has been proposed [6, 26]. An approximate symmetry of order *k* is defined as a solution of the equation

$$\begin{aligned} \partial_t \tilde{S}_n &= \frac{\partial}{\partial v} F\left[\psi + v\tilde{S}_n\right]\Big|_{v=0} + O\left(\varepsilon^{k+1}\right) \\ \tilde{S}_n[\psi] &= S_n[\psi] + \sum_{j=1}^{k} \varepsilon^j S_n^{(j)}[\psi] + O\left(\varepsilon^{k+1}\right) \end{aligned} \quad . \tag{3.9}$$

In Eq. (3.9), *F* is defined by Eq. (3.1) and $S_n$ are the symmetries of the unperturbed NLS equation, defined by Eq. (2.7). To be able to solve for the correction terms $S_n^{(j)}$, the coefficients of the perturbation must obey the same constraints required for regaining the integrability of the NF. For instance, to be able to have (the first non-trivial) $\tilde{S}_3$ of order 2, Eq. (3.7) must be obeyed [6, 26].

The only exception is when the zero-order solution, $\varphi$, is a single soliton. As the NF can be constructed for that case, its symmetries are the symmetries of the unperturbed NLS equation.

### 3.3 Loss of conservation laws

In general, it is impossible to construct conserved quantities for the perturbed NLS equation. As a substitute, the concept of approximate conservation laws has been proposed [26]. An approximate integral of motion of order $k$ is defined as

$$\begin{aligned}
&\tilde{I}_n = \int_{-\infty}^{+\infty} \tilde{\rho}_n[\psi]dx \\
&\frac{d}{dt}\tilde{I}_n = O\left(\varepsilon^{k+1}\right) \\
&\tilde{\rho}_n[\psi] = \rho_n[\psi] + \sum_{j=1}^{k} \varepsilon^j \rho_n^{(j)}[\psi] + O\left(\varepsilon^{k+1}\right)
\end{aligned} \quad . \tag{3.10}$$

In Eq. (3.10), $\rho_n$ are the densities that correspond to the unperturbed equation (see Eq. (2.9)). To be able to solve for the correction terms $\rho_n^{(j)}$, the coefficients of the perturbation must obey the same constraints required for regaining the integrability of the NF [26]. For instance, to be able to have $\tilde{I}_2$ of order 2, Eq. (3.7) must be obeyed. The exception is the case in which the zero-order solution, $\varphi$, is a single soliton. As the NF can be constructed for that case, its solutions obey the same conservation laws as those of the unperturbed NLS equation.

In addition to the inability to construct approximate constants of motion for solutions of the perturbed equation, the zero-order term does not obey any of the conservation laws obeyed by the solution of the unperturbed equation. For any of the quantities defined by Eq. (2.8), one has:

$$\frac{d}{dt}I_n[\varphi] = O\left(\varepsilon^2\right) \quad . \tag{3.11}$$

Consider, for example, $I_0$. Employing Eq. (3.8) for the zero-order approximation, one has:

$$\begin{aligned}
\frac{d}{dt}I_0[\varphi] &= \int_{-\infty}^{+\infty} \left(\varphi_t \varphi^* + \varphi \varphi^*{}_t\right)dx \\
&= \int_{-\infty}^{+\infty} \left(S_2[\varphi]\varphi^* + \varphi S_2[\varphi]^*\right)dx + \varepsilon\alpha_1 \int_{-\infty}^{+\infty} \left(S_3[\varphi]\varphi^* + \varphi S_3[\varphi]^*\right)dx \\
&+ \varepsilon^2 \left\{\beta_1 \int_{-\infty}^{+\infty} \left(S_4[\varphi]\varphi^* + \varphi S_4[\varphi]^*\right)dx + \Gamma \int_{-\infty}^{+\infty} \left(\tilde{R}[\varphi]\varphi^* + \varphi \tilde{R}[\varphi]^*\right)dx\right\}
\end{aligned} \quad . \tag{3.12}$$

Except for the last one, each integrand in Eq. (3.12) is a complete differential. (This can be seen by either direct substitution of the expressions for the symmetries (Eq. (2.7)), or by proof by in-duction, exploiting the recursion relation obeyed by the symmetries [2-5].) Hence, for solutions that vanish at $x = \pm\infty$ , the first three integrals vanish. Thus, for a solution that vanishes at $x = \pm\infty$, Eq. (3.12) becomes

$$\frac{d}{dt} I_0[\varphi] = \varepsilon^2\, \Gamma \int_{-\infty}^{+\infty} \left( \tilde{R}[\varphi]\varphi^* + \varphi \tilde{R}[\varphi]^* \right) dx \quad . \tag{3.13}$$

In general, this integral does not vanish.

Depending on the structure of $\tilde{R}[\varphi]$, this $O(\varepsilon^2)$ violation of the conservation laws may lead to unbounded behavior in the zero-order approximation, $\varphi$ [26].

## 4. New expansion algorithm

### 4.1 Algebraic vs. dynamical obstacles and "canonical" dynamical obstacles

The observation that the same constraints on combinations of coefficients, such as in Eq. (3.7), are required for the three properties that are included in the broad definition of integrability (existence of closed form expressions for single and multiple-soliton solutions, an infinite hierarchy of symmetries and an infinite sequence of conservation laws) is not fully understood. These combinations of coefficients do not vanish for most physical systems. When they do not vanish, they constitute *algebraic* obstacles to integrability. One must, therefore, focus attention on the dynamical quantities that multiply these combinations of coefficients. These are the *dynamical obstacles*, the terms that are unaccounted for in the standard analysis. However, whereas the algebraic obstacles are unique, the dynamical ones are not. This non-uniqueness is a consequence of the freedom in the expansion, as explained in the discussion following Eq. (3.8).

A unique choice of the obstacles can be made by taking into account the observation that none of the problems delineated above emerges when the zero-order solution, $\varphi$, is a single-soliton. Consider, for example, the quantity $I_0$, when it is computed for a solution of the perturbed NLS equation, Eq. (3.1) If a solution, $\psi$, that vanishes at $x = \pm\infty$ exists, one obtains

$$\begin{gathered} \frac{d}{dt} I_0[\psi] = \varepsilon^2 \gamma \int_{-\infty}^{+\infty} i \left\{ \left( |\psi|^2 \psi^* \psi_{xx} - \left(\psi^* \psi_x\right)^2 \right) - \left( |\psi|^2 \psi \psi^*{}_{xx} - \left(\psi \psi^*{}_x\right)^2 \right) \right\} dx \\ \left( \gamma = \left( \frac{\beta_2 - \beta_3 - \beta_4}{2} \right) \right) \end{gathered} \quad . \tag{4.1}$$

For an $O(\varepsilon^2)$ estimate, one is allowed to replace $\psi$ in Eq. (4.1) by the zero-order approximation, $\varphi$. The solution of Eq. (3.1) cannot be a single NLS soliton. However, let us see what happens if one assumes that $\varphi$ is a single NLS soliton. Substituting in Eq. (4.1) the single soliton solution, Eq. (2.2), one finds that the integrand vanishes explicitly. Thus, $I_0[\psi]$ is conserved at least through $O(\varepsilon^2)$. For a general zero-order solution, the integrand in Eq. (4.1) need not vanish, reflecting the effect of the $O(\varepsilon^2)$ obstacle to integrability. Let us recall that the effect of the obstacle is encountered also in the attempt to implement the other two properties incorporated in the concept of integrability (an infinite family of soliton solutions expressible in terms of closed form expressions and an infinite hierarchy of symmetries) for a solution, which is not a single soliton. Naturally, one searches for the dynamic obstacle as a part of the second-order perturbation that vanishes explicitly for a single soliton, but not for other solutions. It is a simple exercise to show that the second-order perturbation in Eq. (3.1) contains a term of the form $\gamma R$, where $R$ is given by

$$R[\psi] = i\left(|\psi|^2 \psi_{xx} - \psi^* (\psi_x)^2 + |\psi|^4 \psi\right) \ , \tag{4.2}$$

and that the integrand in Eq. (4.1) can be written in terms of $R$ as

$$i\left\{\left(|\psi|^2 \psi^* \psi_{xx} - \left(\psi^* \psi_x\right)^2\right) - \left(|\psi|^2 \psi \psi^*_{xx} - \left(\psi \psi^*_x\right)^2\right)\right\} = R\psi^* + R^* \psi \ . \tag{4.3}$$

Substituting in Eq. (4.2) the single soliton solution, one finds that the $R$ vanishes explicitly. This is why the integrand in Eq. (4.1) vanishes. Note that if the coefficient $\gamma$ were to vanish, the violation of the conservation laws would not have been discovered by this imprecise argument. Repetition of this exercise for the next few quantities that are conserved in the unperturbed case reveals that it is the same dynamical obstacle, $R$, which is responsible for the loss of their conservation.

In Eq. (4.1), one is dealing with a bilinear form that involves $\psi_t$ and $\psi^*_t$. However, for the other two components of integrability (an infinite family of multiple-soliton solutions and an infinite hierarchy of symmetries) one deals with $\psi_t$ and $\psi^*_t$ directly, and finds that no obstacles emerge in the single-soliton case. This calls for exploitation of the freedom in the expansion, to obtain a dynamical obstacle that vanishes explicitly in the case of a single soliton, and, hence, reflects the net effect of the difference between single- and multiple-soliton solutions. Starting with the single-soliton values of the coefficients in $\varphi^{(2)}$ given in Eq. (A.1), the value of $B$ that leads to a unique identification of the dynamical obstacle as $R$ of Eq. (4.2), is given in Eq. (A.2). Substitut-ing Eqs. (3.2), (3.5), (3.6), (3,7), (A.1) and (A.2) in Eq. (3.1), and employing the NF, Eq. (3.3), the term, which is left unaccounted for in the second-order calculation, is found to be

$$\tilde{R}[\varphi] = \tfrac{1}{6} R[\varphi] \quad . \tag{4.4}$$

This unique form of the obstacle can be written in terms of symmetries of the unperturbed equation as

$$R[\varphi] = S_2[\varphi] \left| S_0[\varphi] \right|^2 + S_1[\varphi]^2 S_0^*[\varphi] - \left| S_0[\varphi] \right|^4 S_0[\varphi] \quad . \tag{4.5}$$

To summarize, the reason why obstacles to integrability do not emerge when the zero-order approximation is a single NLS soliton is that the term that is responsible for spoiling integrability vanishes explicitly in that case.

**4.2 Shifting loss of integrability to NIT**

With the assumption that the higher-order corrections are differential polynomials in $\varphi$ [6, 25, 26], the unaccounted for quantity, $\Gamma \tilde{R}[\varphi]$, must be included in the second-order contribution to the NF, as in Eq. (3.8), spoiling its integrability. To avoid this state of affairs, one must allow the second-order correction, $\varphi^{(2)}$, to have a non-polynomial dependence on $\varphi$. The simplest way to realize this idea is by adding to $\varphi^{(2)}$ of Eq. (3.5) a term that is explicitly $t$- and $x$-dependent, modifying Eq. (3.5) into

$$\varphi^{(2)} = A\varphi_{xx} + B\varphi^2 \varphi^* + C\,p\varphi + D\,q\varphi_x + E\,q^2 \varphi + \xi(t,x) \quad . \tag{4.6}$$

Substituting Eqs. (3.2), (3.5) (modifying $\varphi^{(2)}$ according to Eq. (4.6)), (3.6), (A.1) and (A.2) in Eq. (3.1), and employing the NF, Eq. (3.3), the equation obeyed by $\xi(t,x)$ is found to be

$$\partial_t \xi = i\partial_x^{\,2} \xi + 2i\left(2|\varphi|^2 \xi + \varphi^2 \xi^*\right) + \Gamma \tilde{R}[\varphi] \quad . \tag{4.7}$$

Thus, allowing for non-polynomial dependence in the $O(\varepsilon^2)$ correction to the solution, it is possible to account for the dynamical obstacle to integrability even when the algebraic obstacle does not vanish. The NF remains Eq. (3.3) because there is no need to include the obstacle in it. This NF is integrable and its solutions are the single- and multiple-soliton solutions of the unperturbed equation, with the parameters of each soliton modified by Eq. (3.4). These solutions obey the same infinite sequence of conservation laws obeyed by the unperturbed solutions, and the NF has the same hierarchy of symmetries as the unperturbed equation. The price paid is that the second-order correction to the solution contains a non-polynomial dependence on $\varphi$, which may have to be computed numerically.

Detailed calculations show that the algorithm proposed here does not overcome the loss of the ability to construct approximate symmetries and approximate integrals of motion, as defined in Sections 3.2 and 3.3, respectively, for Eq. (3.1) and its full solution, $\psi$. The algebraic obstacles to integrability are not eliminated as far as the construction of these quantities is concerned. Only the NF and its solution, the zero-order approximation, satisfy the three aspects of integrability.

### 4.3 Effect of "canonical" obstacle on multiple-soliton solution

The canonical form of the dynamical obstacle, given by Eqs. (4.2) and (4.4), is particularly useful when the zero-order approximation, $\varphi$, is a multiple-soliton solution. The reason is that, as $R[\varphi]$ vanishes explicitly for a single-soliton solution, in the multiple-soliton case it vanishes away from the origin in the *x-t* plane, where the solution asymptotes into well-separated single solitons. This is demonstrated here for the two-soliton case. The expression for the two-soliton solution is given by Eqs. (2.4), (2.5) and (2.3). Fig.1 shows the absolute value of a two-soliton solution, and Fig. 2 - the absolute value of the dynamical obstacle $R$ of Eq. (4.2).

To see the effect of the obstacle on the non-polynomial second-order contribution, $\xi(t,x)$, the solution of Eq. (4.7), we rewrite $R$ of Eq. (4.2) as:

$$R[\varphi] = i|\varphi|^2 \varphi\left(|\varphi|^2 + \partial_x\left(\frac{\varphi_x}{\varphi}\right)\right) \quad . \tag{4.8}$$

Using the explicit form of the two-soliton solution, given by Eq. (2.4), (2.3) and (2.5), one finds that the asymptotic behavior of $\varphi(t,x)$ along each soliton is dominated by

$$\varphi \to \begin{cases} e^{i\gamma_1}\left\{O(1) + O\left(e^{-|\beta_2|}\right)\right\} & |\beta_1| = O(1)\,, |\beta_2| \to \infty \\ e^{i\gamma_2}\left\{O(1) + O\left(e^{-|\beta_1|}\right)\right\} & |\beta_1| \to \infty\,, |\beta_2| = O(1) \end{cases} \quad . \tag{4.9}$$

Away from the characteristic lines of both solitons, one finds

$$|\varphi| \propto e^{-Max\left[|\beta_1|,|\beta_2|\right]} \qquad |\beta_1|,|\beta_2| \to \infty \quad . \tag{4.10}$$

The quantity in parentheses in Eq. (4.8) vanishes identically for a single-soliton solution. A detailed study of its structure in the two-soliton case yields that the $O(1)$ contribution contained in $\varphi(t,x)$ is eliminated in that quantity, and that the leading behavior of $R[\varphi]$ along each soliton is:

$$R[\varphi] \propto \begin{cases} e^{i\gamma_1} e^{-|\beta 2|} & |\beta_1| = O(1)\,, |\beta_2| \to \infty \\ e^{i\gamma_2} e^{-|\beta_1|} & |\beta_1| \to \infty\,, |\beta_2| = O(1) \end{cases} \,. \tag{4.11}$$

Away from the characteristic lines of both solitons, one finds

$$|R[\varphi]| \propto e^{-3\,Max\{|\beta_1|,|\beta_2|\}}\, e^{-|\beta_1|-|\beta_2|} \quad |\beta_1|,|\beta_2| \to \infty \quad . \tag{4.1}$$

The exponential decay of $|R[\varphi]|$ is demonstrated in Figs. 3 and 4, where Log($R$) is plotted.

As the driving term in Eq. (4.7), falls off exponentially, $\xi(t,x)$, the non-polynomial second-order term in the NIT (solution of Eq. (4.7)), also falls off exponentially in all directions in the plane. Thus, away from the origin in the $x$-$t$ plane, the second order correction to the solution asymptotes to its part that can be written as a differential polynomial in $\varphi$.

These conclusions do not hold for other forms of the obstacle to integrability. Once the obstacle does not vanish asymptotically along each of the solitons, it has the capacity to generate unbounded behavior in $\xi(t,x)$.

## 5. Concluding remarks

The effect of obstacles to integrability in the perturbed NLS equation can be shifted from the NF to the NIT by allowing the higher-order corrections in the NIT to depend explicitly on $t$ and $x$. The penalty for the loss of integrability is the fact that the NIT ceases to be a sum of differential polynomials in the zero-order approximation. Some contributions to the NIT have a non-polynomial structure and may have to be found numerically. The gain is that the NF is constructed from symmetries only, hence, remains integrable, and the zero-order term retains the multiple-soliton structure of the unperturbed solution.

The "canonical" obstacle generated by our algorithm (see Eq. (4.2)) is expressible in terms of symmetries of the unperturbed equation and vanishes explicitly in the case of a single-soliton solution of the NF. Moreover, as the canonical obstacle decays rapidly away from the interaction region in the multiple-soliton case, it does not generate unbounded behavior in the solution, but only a decaying tail, which emanates from the soliton-interaction region around the origin. In [26] it is suggested that the origin of obstacles to integrability is inelastic interactions. In the unperturbed NLS equation, inelastic interactions amongst the exponential waves, from which a multiple-soliton solution is constructed, are confined to the soliton collision region. With canonical obstacles, the inelasticity remains confined to the same region.

With canonical obstacles, the second order correction to the solution (Eq. (4.6)) tends asymptotically away from the origin to its closed-form, differential-polynomial part. Thus, "asymptotic" integrability acquires a new meaning. The approximate solution is given by a zero-order term, which is determined by an integrable NF (namely, constructed from symmetries only), and an NIT, the structure of which tends to that of the integrable case (when no obstacles exist) asymptotically away from the origin.

The extension to higher orders, while cumbersome, is obvious. Denoting the $n$'th-order term in the NIT by $\varphi^{(n)}$, one exploits the freedom in the determination of the coefficients of monomials in the differential polynomial that the formalism allows to have in $\varphi^{(n)}$, so as to obtain a canonical obstacle to integrability. One then allows $\varphi^{(n)}$ to have a non-polynomial contribution. The driving term in the equation that determines the latter is localized around the origin in the $x$-$t$ plane. It contains the $n$'th-order canonical obstacle, as well as the cumulative effect of (non-polynomial) terms that have been computed (perhaps numerically) in lower orders.

Finally, the results presented here apply also when the higher-order perturbation in Eq. (3.1) are not small, namely, $\varepsilon = O(1)$, provided $|\varepsilon \cdot K| \ll 1$, where $K$ is the largest wave number considered. The reason is that the true expansion parameter is not $\varepsilon$, but $\varepsilon \cdot K$.

31. **Appendix**

Coefficients in $\varphi^{(2)}$ (Eq. (3.5)) in single soliton case

$$\begin{aligned}
A &= \tfrac{9}{8}\alpha_1^{\ 2} + \tfrac{3}{8}\alpha_1\alpha_2 - \tfrac{3}{32}\alpha_2^{\ 2} - \tfrac{3}{8}\alpha_1\alpha_3 + \tfrac{3}{16}\alpha_2\alpha_3 - \tfrac{3}{32}\alpha_3^{\ 2} \\
&\quad + 3\beta_1 - \tfrac{1}{4}\beta_2 - \tfrac{1}{4}\beta_3 - \tfrac{1}{4}\beta_4 + \tfrac{1}{4}\beta_5 \\
C &= -2B + 3\alpha_1^{\ 2} + \alpha_1\alpha_2 - \tfrac{1}{4}\alpha_2^{\ 2} - \tfrac{3}{2}\alpha_1\alpha_3 + \tfrac{5}{12}\alpha_2\alpha_3 - \tfrac{1}{12}\alpha_3^{\ 2} \\
&\quad + 8\beta_1 - \tfrac{1}{3}\beta_2 - \tfrac{2}{3}\beta_3 - \tfrac{1}{3}\beta_4 + \tfrac{1}{3}\beta_5 - \tfrac{1}{3}\beta_6 \\
D &= -2B + \tfrac{15}{2}\alpha_1^{\ 2} + \tfrac{5}{2}\alpha_1\alpha_2 - \tfrac{5}{8}\alpha_2^{\ 2} - \tfrac{9}{4}\alpha_1\alpha_3 + \tfrac{19}{24}\alpha_2\alpha_3 - \tfrac{1}{12}\alpha_3^{\ 2} \\
&\quad + 20\beta_1 - \tfrac{4}{3}\beta_2 - \tfrac{5}{3}\beta_3 - \tfrac{1}{3}\beta_4 + \tfrac{5}{6}\beta_5 - \tfrac{1}{3}\beta_6 \\
E &= -2B + \tfrac{3}{2}\alpha_1^{\ 2} + \tfrac{7}{2}\alpha_1\alpha_2 - \tfrac{5}{8}\alpha_2^{\ 2} - \tfrac{3}{2}\alpha_1\alpha_3 + \tfrac{7}{12}\alpha_2\alpha_3 + \tfrac{1}{12}\alpha_3^{\ 2} \\
&\quad + 14\beta_1 - \tfrac{2}{3}\beta_2 - \tfrac{4}{3}\beta_3 + \tfrac{1}{3}\beta_4 + \tfrac{2}{3}\beta_5 - \tfrac{2}{3}\beta_6
\end{aligned} \quad . \qquad \text{(A.1)}$$

Coefficient $B$ in $\varphi^{(2)}$ (Eq. (3.5)) generating "canonical" obstacle to integrability

$$\begin{aligned}
B &= 3\alpha_1^{\ 2} + \alpha_1\alpha_2 - \tfrac{1}{4}\alpha_2^{\ 2} - \tfrac{3}{4}\alpha_1\alpha_3 + \tfrac{7}{24}\alpha_2\alpha_3 + \tfrac{1}{24}\alpha_3^{\ 2} \\
&\quad + 7\beta_1 - \tfrac{1}{3}\beta_2 - \tfrac{2}{3}\beta_3 + \tfrac{1}{6}\beta_4 + \tfrac{1}{3}\beta_5 - \tfrac{1}{3}\beta_6
\end{aligned} \quad . \qquad \text{(A.2)}$$

Figure captions

Fig. 1 Absolute value of two-soliton solution (Eqs. (2.4), (2.5) & (2.3)); $k_1 = 0.5$, $v_1 = 1$, $k_2 = 0.6$, $v_2 = 1.5$, $x_{1,0} = x_{2,0} = 0$.

Fig. 2 Absolute value of canonical obstacle $R$ (Eq. (4.2)) for two-soliton solution; parameters as in Fig. 1.

Fig. 3 $x$-dependence of $|R|$ (Eq. (4.2)) at $t = 0$ (thick line), 1 (thin line), 2 (long dashes), 4 (short dashes) for two-soliton solution; parameters as in Fig. 1.

Fig. 4 $t$-dependence of $|R|$ (Eq. (4.2)) at $x = 0$ (thick line), 1 (thin line), 2 (long dashes), 4 (short dashes) for two-soliton solution; parameters as in Fig. 1.

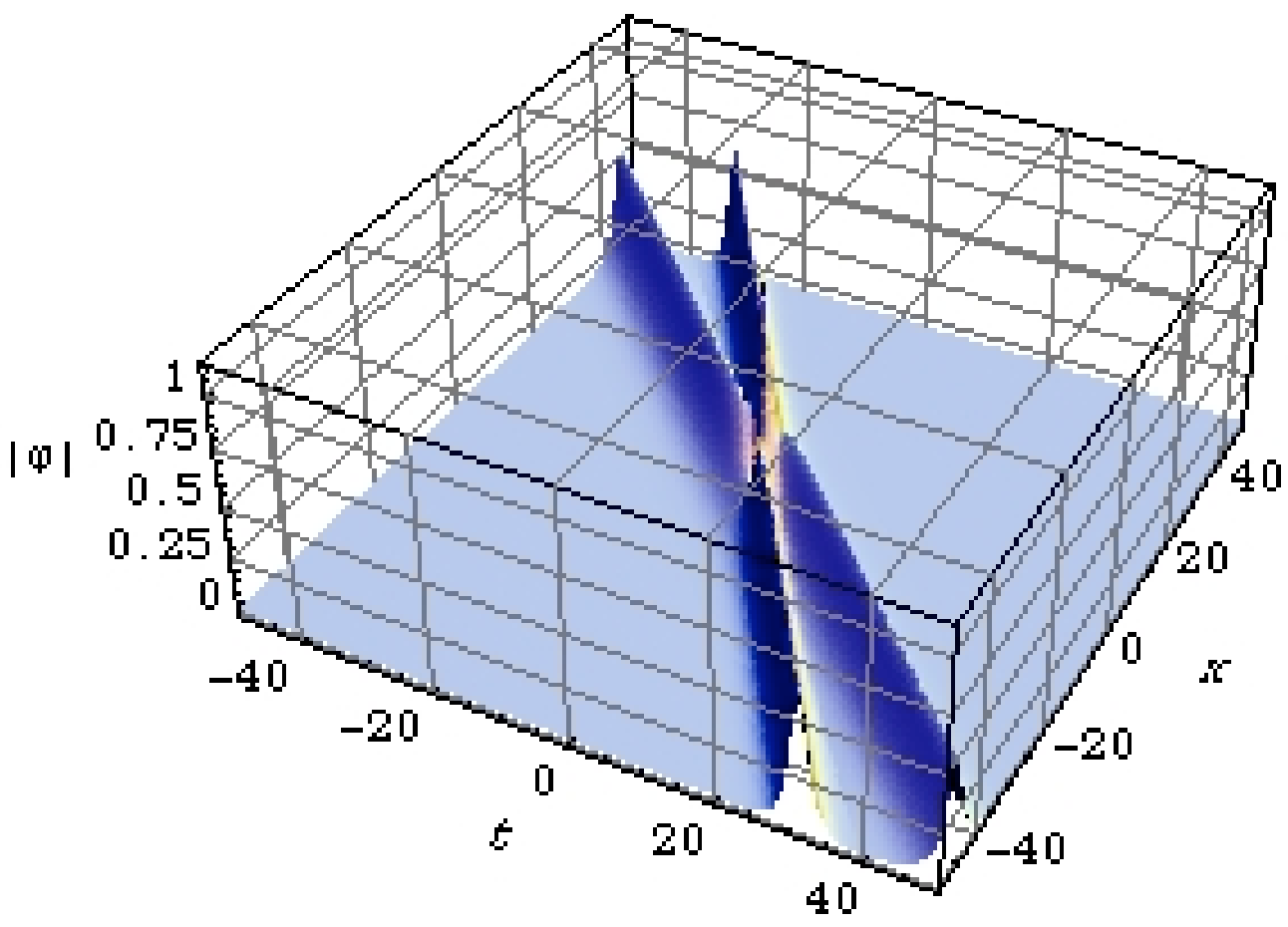

1
0.75
|φ|
0.5
0.25
0
-40
-20
0
t
20
40
40
20
0
x
-20
-40

Fig. 1

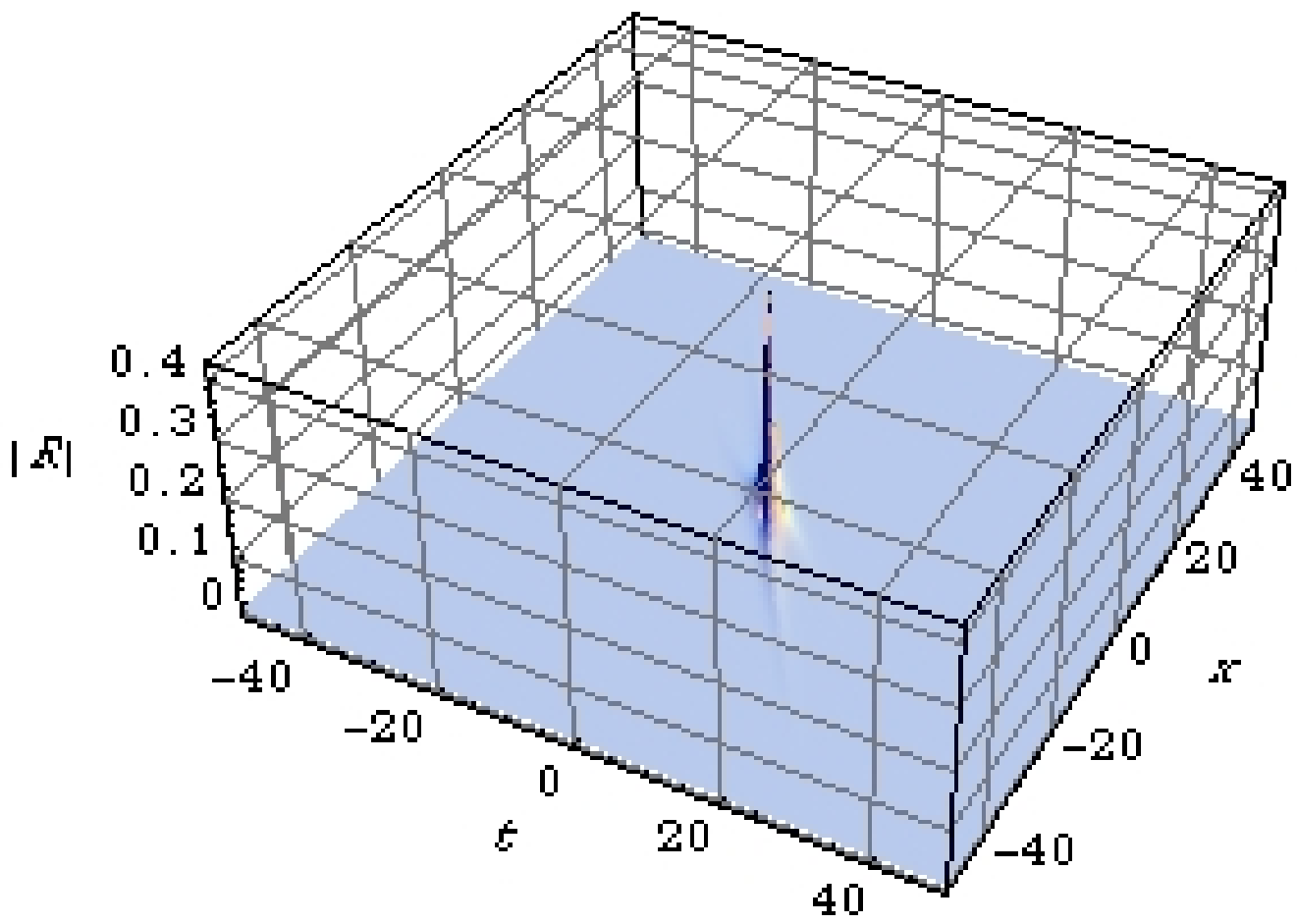

0.4
0.3
|E|
0.2
0.1
0
-40
-20
0
t
20
40
40
20
0
x
-20
-40

Fig. 2

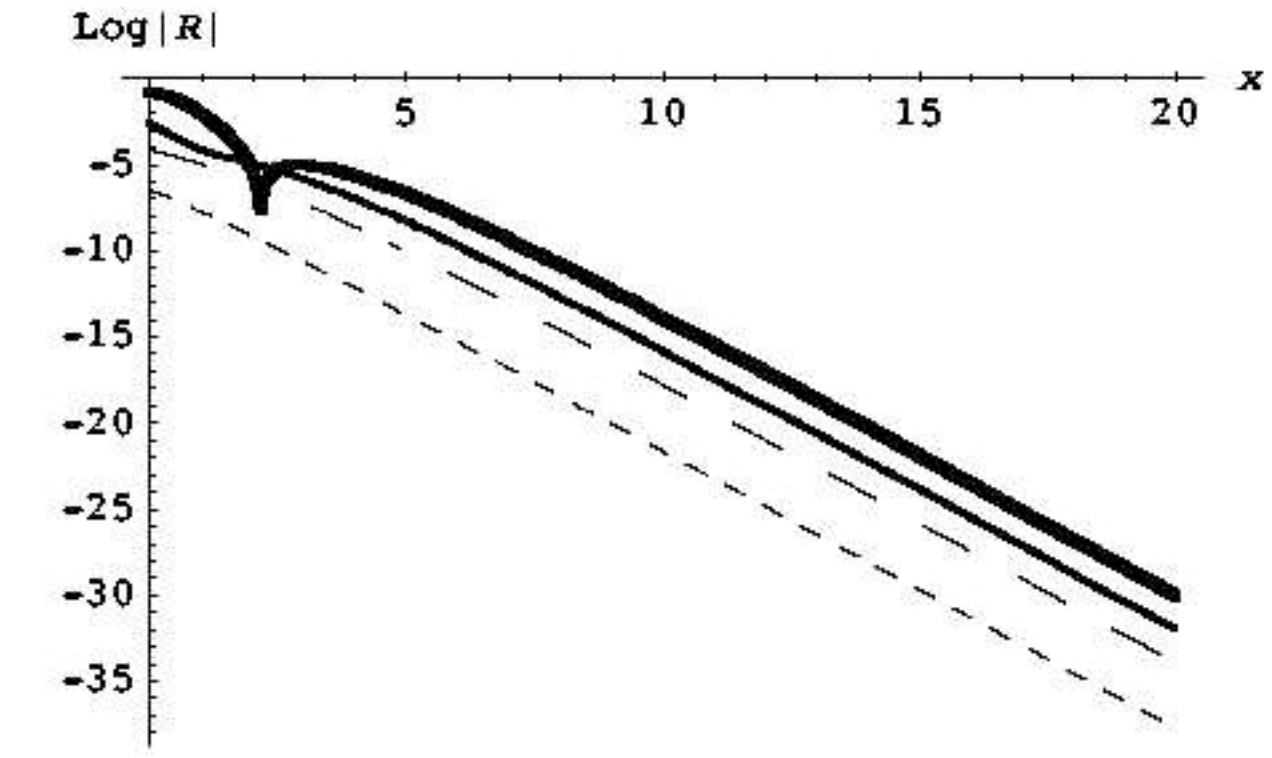

Fig. 3

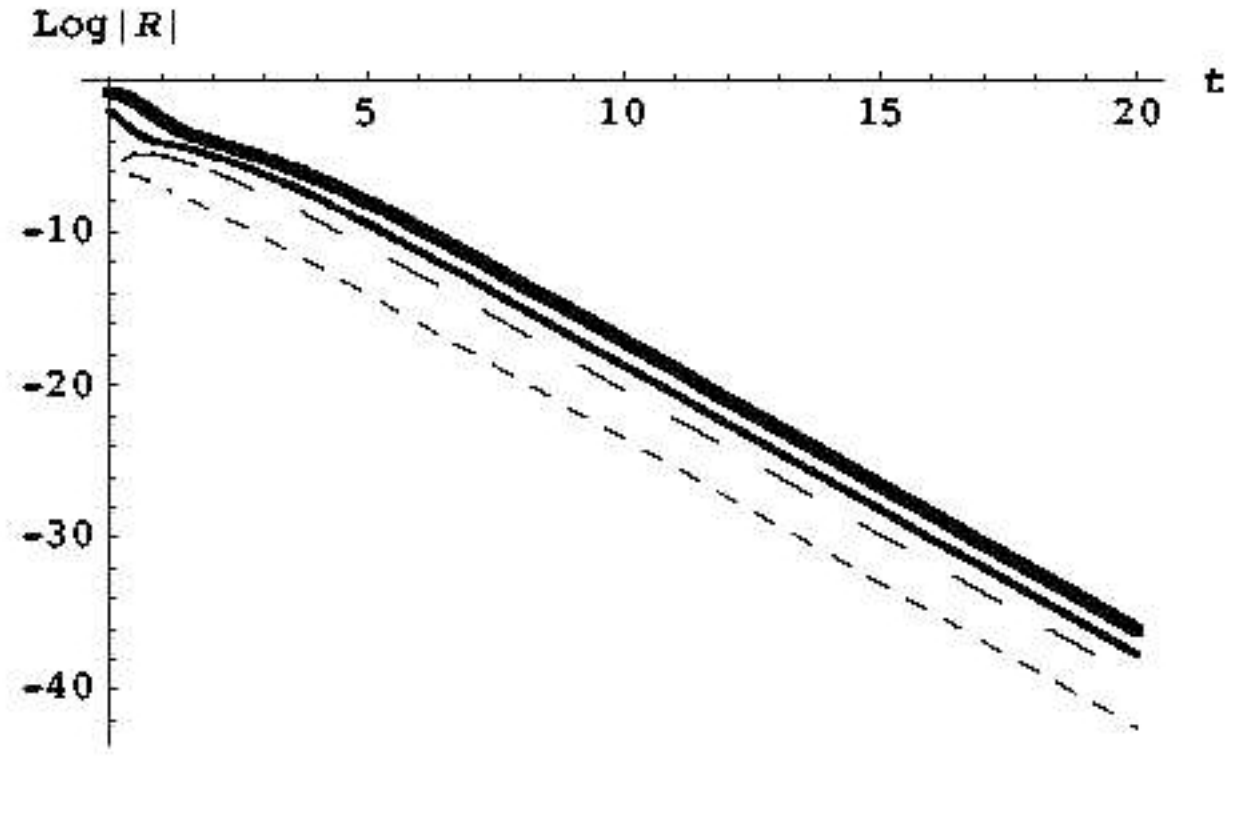

Fig. 4